\documentclass[aps, prd, reprint, amsfonts,
 amssymb, amsmath, showpacs, letterpaper, nofootinbib, groupedaddress]{revtex4-1}

\usepackage{braket, bm}

\usepackage{color,graphicx}

\usepackage{ulem}
\usepackage{comment}
\usepackage{amssymb, amsmath}

\usepackage{slashed}

\usepackage[caption=false]{subfig}

\newcommand{\be}{\begin{equation}}
\newcommand{\ee}{\end{equation}}
\newcommand{\lagr}{{\cal L}}

\DeclareMathOperator\arctanh{arctanh}
\DeclareMathOperator\am{am}
\DeclareMathOperator\sn{sn}

\DeclareMathOperator\cd{cd}
\DeclareMathOperator\nd{nd}
\DeclareMathOperator\sd{sd}
\DeclareMathOperator\nc{nc}

\begin{document}

\title{Particle level screening of scalar forces in 1+1 dimensions}

\author{Clare Burrage}
\email{clare.burrage@nottingham.ac.uk}

\author{Benjamin Elder}
\email{benjamin.elder@nottingham.ac.uk}

\author{Peter Millington}
\email{p.millington@nottingham.ac.uk}
\affiliation{School of Physics and Astronomy, University of Nottingham, Nottingham NG7 2RD, United Kingdom}
%




\begin{abstract}
 We investigate how non-linear scalar field theories respond to point sources.  Taking the symmetron as a specific example of such a theory, we solve the non-linear equation of motion in one spatial dimension for (i) an isolated point source and (ii) two identical point sources with arbitrary separation.  We find that the mass of a single point source can be screened by the symmetron field, provided that its mass is above a critical value.  We find that two point sources behave as independent, isolated sources when the separation between them is large, but, when their separation is smaller than the symmetron's Compton wavelength, they behave much like a single point source with the same total mass.  Finally, we explore closely related behavior in a toy Higgs-Yukawa model, and find indications that the maximum fermion mass that can be generated consistently via a Yukawa coupling to the Higgs in 1+1 dimensions is roughly the mass of the Higgs itself, with potentially intriguing implications for the hierarchy problem. \end{abstract}

\maketitle
\section{Introduction}
\label{intro}

Scalar field theories with non-linear equations of motion are increasingly common in modern physical models.  Their non-linearities can give rise to rich phenomenological behavior and have found applications ranging from fundamental particle physics and cosmology to superconductivity and superfluids.  Perhaps the best-known example is the Higgs mechanism~\cite{Englert:1964et, Higgs:1964pj, Guralnik:1964eu}, where a quartic potential gives rise to spontaneous symmetry breaking at low energy.  This mechanism generates fermion masses via Yukawa couplings without breaking gauge symmetry in the Standard Model (SM).

Another fruitful application of non-linear scalar field theories is to cosmology.  Scalar fields are regularly employed in models of dark energy~\cite{Joyce:2014kja}, dark matter~\cite{Bertone:2004pz}, and inflation~\cite{Akrami:2018odb}.  These theories often include explicit couplings to matter or generate them quantum-mechanically.  However, those couplings are strongly constrained by fifth force tests~\cite{Adelberger:2003zx}, unless the theory contains a {\it screening mechanism} that dynamically suppresses the force mediated by the scalar field.  The symmetron~\cite{Hinterbichler:2010es}, chameleon~\cite{Khoury:2003aq}, and Vainshtein~\cite{Vainshtein:1972sx,Deffayet:2001uk,Nicolis:2008in} mechanisms are all well-known examples of this behavior.  See Ref.~\cite{Joyce:2014kja} for a review of these theories.  All of these theories contain some form of non-linearity in their equation of motion which allows them to screen.

Unfortunately, the phenomenological richness of screened theories usually comes at the cost of a complicated equation of motion that is difficult to solve in general.  Analytic studies typically rely on linearizing the theory in different regimes and then building up piecewise solutions to approximate the full non-linear solution.
Another common approximation is to treat the source distribution as smooth, even though realistic sources have significant substructure at atomic and molecular scales.

Although linearization is a powerful tool, there is no substitute to a full solution of the non-linear equation of motion.  Such solutions can be useful for verifying the qualitative behavior found in the linear treatment.  They can also do much to reveal non-linear phenomena that may  have otherwise  been missed.

It is also of interest to ask how the story changes when we couple to individual particles, i.e.~highly localized sources of matter.  One could hope to learn when it is justified to  treat the matter as a smooth distribution, and when the matter must be treated as point-like.

In this paper we explore the dynamics of a non-linear scalar field coupled to point sources.  We take the symmetron and chameleon as the archetypal theories, which have quadratic and linear couplings to matter, respectively.  We focus mainly on the symmetron field, and derive solutions to the non-linear equation of motion in the presence of point-like sources of matter.  We retreat to one spatial dimension, where it is possible to proceed analytically, and take the density of the sources to be approximated by Dirac delta functions. As we will show, this approximation is valid so long as the Compton wavelength of the screened field is much larger than that of the sources. In this way, we do not need to address the problem of localization in quantum field theory.

First, we construct vacuum solutions around a single point source of matter.  Our main finding here is that the point source can be screened by the symmetron field.  This is somewhat unexpected, given that the traditional requirements for symmetron screening are that a source be both sufficiently dense and large.  For the point source, the only requirement is that its mass be larger than a critical value.

We also study vacuum solutions around two identical point sources of mass $m$.  We find that, when the distance between the sources is large, the solution in the vicinity of each source mimics the solution around a single source of mass $m$.  However, when their separation is less than a Compton wavelength of the symmetron, the external solution resembles that of a single point source with total mass $2m$, and, importantly, the field is roughly constant between the sources.

We then briefly discuss the electrostatic analogy for the symmetron theory \cite{JonesSmith:2011tn,Pourhasan:2011sm}, which has proven to be a powerful tool for understanding symmetron and chameleon dynamics and screening.  It was recently argued, within the electrostatic approximation, that the symmetron force between a small, unscreened point charge and a large, screened object could become {\it repulsive} when the separation between them is sufficiently small \cite{Ogden:2017xeo}.  We show that in one spatial dimension the symmetron force resulting from the full non-linear theory is always attractive under quite general assumptions.

Finally, we return to solutions with a single source, but for theories with a linear coupling to matter.  We show that chameleons, thanks to their linear matter coupling and unbounded self-interaction potential, do not admit the point-particle screening behavior we found for the symmetron.  We also explore a toy Higgs model, where arbitrarily large couplings to matter lead to a breakdown in the theory.  Consequently, in this model, we find indications that one cannot consistently generate (non-relativistic) fermion masses larger than roughly that of the Higgs mass itself in 1+1 dimensions. 

Our analytic approach follows that of Refs.~\cite{Upadhye:2012rc,Brax:2017hna}, which solved for the chameleon and symmetron equations of motion in effectively one-dimensional planar systems.  Numerical solutions of the chameleon field in a radially symmetric system of spherical matter shells were obtained in Ref.~\cite{Nakamura:2018gxf}.  Non-linear solutions to theories with Vainhstein screening were also explored in Ref.~\cite{Bloomfield:2014zfa}.

This paper is organized as follows.  Section \ref{sec:symmetron-basics} gives an overview of the symmetron theory and the usual route to understanding symmetron screening. Sections \ref{sec:SP-soln}-\ref{sec:double-identical} solve for the field around a single point particle, a top-hat distribution, and a two-particle system, respectively.  Section \ref{electrostatics} discusses the electrostatic analogy of the symmetron.  Section \ref{sec:linear-coupling} explores how the story changes when the coupling to matter is linear rather than quadratic, as is the case with chameleons, and Section \ref{sec:higgs} is devoted to a discussion of related behavior in Higgs-Yukawa theories.

\section{Symmetron overview}
\label{sec:symmetron-basics}
The symmetron is a modification of gravity with a scalar degree of freedom in addition to the usual metric tensor \cite{Hinterbichler:2010es,Hinterbichler:2011ca} (see Refs.~\cite{Dehnen:1992rr,Gessner:1992flm,Damour:1994zq,Pietroni:2005pv,Olive:2007aj,Brax:2010gi} for earlier related work.)   There are two important ingredients: (1) a Higgs-like potential, leading to a non-zero vacuum expectation value (VEV) in regions of low density, and a symmetric phase where the scalar field goes to zero in dense regions, and (2) a matter coupling that scales with the local field value. These properties allow the symmetron to mediate a scalar fifth force, whose strength depends on the environment.

Originally introduced as a candidate to drive cosmic acceleration, symmetrons have also been used to reproduce galactic rotation curves without dark matter \cite{Burrage:2016yjm,OHare:2018ayv}.  Evidence of symmetrons is actively being sought via cosmological tests, astrophysical probes, and experiments in the laboratory \cite{Upadhye:2012rc,Burrage:2016rkv,Brax:2016wjk,Llinares:2018mzl,Brax:2017hna,Brax:2017xho,Brax:2018zfb}. For a recent review of constraints, see Ref.~\cite{Burrage:2017qrf}.

The canonical example of a symmetron theory has a Lagrangian\footnote{We work in flat space and use the mostly-plus sign convention for the metric.}
\be
\lagr = - \frac{1}{2} (\partial \phi)^2 - \frac{1}{2} \left( \frac{- T^\mathrm{m}}{M^2} - \mu^2 \right) \phi^2 - \frac{1}{4} \lambda \phi^4~,
\label{full-lagr}
\ee
 where $T^\mathrm{m}$ is the trace of the energy momentum tensor for matter fields.

For static field configurations and non-relativistic matter, $T^{\rm m}=-\rho$, and the equation of motion becomes
\be
\vec \nabla^2 \phi = \left( \frac{\rho}{M^2} - \mu^2 \right) \phi + \lambda \phi^3~.
\label{full-eom}
\ee
In a perfect vacuum, the energy density $\rho = 0$, and there is a non-zero VEV $\phi_\infty \equiv \pm \mu / \sqrt \lambda$ which spontaneously breaks the $\mathbb{Z}_2$ symmetry of the action.  For sufficiently large $\rho > \mu^2 M^2$, the VEV becomes zero, and the symmetry is restored.

Although theories that couple universally to the matter energy-momentum tensor are not renormalizable, the symmetron mechanism can also be realized through the Coleman-Weinberg mechanism~\cite{Coleman:1973jx} of dimensional transmutation, exploiting the spontaneous breaking of scale symmetry by radiative effects~\cite{Burrage:2016xzz}.

The local matter density appears in Eq.~\eqref{full-lagr} because matter fields move on geodesics of the so-called Jordan-frame metric $g^\mathrm{JF}_{\mu\nu}=(1+\phi^2/M^2) g^{\rm EF}_{\mu\nu}$, where EF indicates the Einstein-frame metric, see Ref.~\cite{Hinterbichler:2010es} for a derivation.  This gives rise to a scalar fifth force for the matter fields, given by
\be
\vec F = - \frac{\phi}{M^2} \vec \nabla \phi~,
\label{test-force}
\ee
for a unit test mass.

Note that the quadratic coupling to matter implies that the strength of the fifth force scales with the ambient field value.  In dense regions, the scalar field is driven to zero and its matter coupling vanishes, suppressing the fifth force.  This effect occurs inside large, dense objects, such that only the matter near the surface of the object is coupled to the scalar field, and only a small fraction of the total mass of the object contributes to the net scalar force, i.e. some of the mass of the object is screened.

The usual route towards understanding symmetron screening is the following: imagine a sphere of density $\rho$ and radius $R$ surrounded by vacuum. The behavior of the field is governed by an effective potential
\begin{equation}
V_{\rm eff}(\phi)= \frac{1}{2}\left(\frac{\rho \Theta(R-r)}{M^2}-\mu^2\right)\phi^2+\frac{\lambda}{4}\phi^4~.
\end{equation}
If $\rho> M^2 \mu^2$ then the effective potential is minimized inside the source by
\begin{equation}
\phi_{\rm in}= 0~,
\end{equation}
while outside the source the minimum of the potential is 
\begin{equation}
\phi_{\infty}= \frac{\mu}{\sqrt{\lambda}}~.
\end{equation}
The corresponding masses of the  field around these minima are $m_{\rm in} = \sqrt{(\rho/M^2)-\mu^2}$ and $m_{\infty}= \sqrt{2}\mu$.

To proceed the following assumptions are made: (i) we approximate the effective potential as quadratic both inside and outside the source, (ii) we impose that the field and its first derivative are continuous at the surface of the source, and (iii) we require that the field decays to its VEV at blue spatial infinity and is regular at the origin.  We then find the approximate solution, up to factors of order $(m_\mathrm{in} R)^{-1}$,
\begin{equation}
\phi(r) = 
\left\{
\begin{array}{lc}
0 & 0\leq r<R~,\\
\phi_{\infty} -\frac{R\phi_{\infty}}{r}e^{-m_{\infty}r} & R<r~,
\end{array}
\right.
\label{piecewise-linear}
\end{equation}
in the regime of interest where $m_{\rm in }R>1$ and $m_{\infty}R<1$.

From Eq.~\eqref{test-force} we see that, within a Compton wavelength of the scalar field in vacuum, the ratio of the symmetron mediated force to the Newtonian gravitational potential sourced by the same mass is 
\be
\frac{F_{\phi}}{F_N}= \frac{8\pi M_\mathrm{Pl}^2}{M^2}\frac{\phi_{\infty}^2R}{M_S}\left(1-\frac{R}{r}\right)=\frac{\phi_{\infty}^2}{M^2 \Phi(R)}\left(1-\frac{R}{r}\right)~,
\ee
where $M_S= 4\pi R^3 \rho/3$ is the mass of the source, $\Phi(R)= M_S/8\pi M_\mathrm{Pl}^2 R$ is the Newtonian potential at the surface of the source, and $M_\mathrm{Pl} \equiv (8 \pi G)^{-1/2}$ is the reduced Planck mass. 

  Note that the force depends only on $R$, and not on the density of the source $\rho$.  Increasing $\rho$, and therefore the total mass of the object, does not result in a stronger force so long as $R$ remains the same.  In other words, the scalar force is suppressed compared to gravity as long as the depth of the symmetron scalar potential is shallower than the corresponding Newtonian potential well as $\phi_{\infty}^2/M^2 < \Phi(R)$.   Some of the mass of this object is therefore screened.
  
  If we had instead made the assumption that $\rho < M^2 \mu^2$, or that $m_{\rm in} R < 1$, we would have found that the force scales linearly with the mass of the object: sufficiently small and light objects are unscreened.

\section{Single point source solution}
\label{sec:SP-soln}
Here, we solve exactly for the symmetron field in one spatial dimension around a point particle of mass $m$ located at a point $x_1>0$.  We assume, without loss of generality, that the density is zero elsewhere.\footnote{As long as the ambient matter density is uniform and less than the symmetry-breaking density $\rho_\mathrm{amb} < \mu^2 M^2$, it may be absorbed into a redefinition of $\mu$.}

We eliminate $\lambda$ from the equation of motion by introducing a new field variable $\varphi \equiv \phi / \phi_\infty$.   We also absorb $\mu$ by defining the dimensionless coordinate $\hat x = \mu x$.

The symmetron Lagrangian, up to an overall constant factor $\mu^2 \phi_\infty^2$, is now
\be
\hat \lagr = - \frac{1}{2} (\hat \partial \varphi)^2 - \frac{1}{2} \left( \frac{\rho}{\mu ^2 M^2} - 1 \right) \varphi^2 - \frac{1}{4} \varphi^4~,
\ee
 where the derivative $\hat \partial$ indicates differentiation with respect to $\hat x$ and $\hat t= \mu t$.  Note that we are working in one spatial dimension, in which case $\phi$ and $M$ are dimensionless, $\mu$ is dimension 1, and $\rho$ and $\lambda$ are dimension 2.

This dimensionless Lagrangian describes a scalar field with a self-interaction potential
\be
V(\varphi) = -\frac{1}{2} \varphi^2 + \frac{1}{4} \varphi^4~.
\ee
The field obeys a static equation of motion
\be
\varphi'' = - \varphi \left(1 - \varphi^2 \right) + \frac{1}{\mu^2 M^2} \varphi \rho~,
\label{varphi-eom}
\ee
where the prime denotes a derivative with respect to $\hat x$.

We solve Eq.~\eqref{varphi-eom} in the same way as Refs.~\cite{Upadhye:2012rc,Brax:2017hna} but with a central point source\footnote{We define the source as $\rho(x) = m \delta(x - x_1)$, but when we move to the dimensionless coordinate $\hat x$ it picks up a factor of $\mu$, such that $\rho(\hat x) =\mu m \delta(\hat x - \hat x_1)$.} $\rho(\hat x) = \mu m \delta( \hat x - \hat x_1)$, subject to the boundary conditions $\varphi \to 1$ as $\hat x \to \pm \infty$.

We first focus on the region $\hat x < \hat x_1$.  Rewriting $\varphi'' = \frac{1}{2}\frac{\rm d}{\rm d \varphi} (\varphi'^2)$, and integrating from $\varphi(-\infty)$ to $\varphi(\hat x)$, we have
\be
 \varphi'^2 = 2 \left( V|_{\hat x} - V|_{-\infty} \right)~,
\ee
where we have assumed that the field gradient vanishes at $\hat x = \pm \infty$.  We take the positive root, plug in the definition of $V$, separate variables, and then integrate once again, this time between some $\hat x < \hat x_1$ and $\hat x_1$:
\be
 \int_{\hat x}^{\hat x_1^-} \rm d \hat x' = \int_{\varphi(\hat x)}^{\varphi(\hat x_1^-)} \frac{\rm d \varphi}{\sqrt{\frac{1}{2} - \varphi^2 + \frac{1}{2} \varphi^4}}~.
 \label{implicit-intermediate}
\ee
The denominator on the RHS may be rewritten as $\sqrt{\frac{1}{2} - \varphi^2 + \frac{1}{2} \varphi^4} = \frac{1}{\sqrt 2} (1 - \varphi^2)$.  The resulting integral is straightforward:
\be
\int \frac{{\rm d} y}{1 - y^2 } = \arctanh y + \mathrm{const.}~,
\ee
leading to the implicit relation
\be
\hat x_1 - \hat x = \sqrt 2 \left( \arctanh \varphi_1 -  \arctanh \varphi(\hat x) \right)~,
\ee
where $\varphi_1 \equiv \varphi(\hat x_1)$ is the field value at the point source.
Inverting, we find
\be
\varphi(\hat x)  = \tanh \left( \frac{1}{\sqrt 2} | \hat x - \hat x_1 | + \arctanh \varphi_1   \right)~.
\label{single-particle-soln}
\ee
It is easily checked that this is the solution for $\hat x > \hat x_1$ as well.

All that remains is to determine the integration constant $\varphi_1$, which gives the field value at the source.  We do so by integrating Eq.~\eqref{varphi-eom} over an infinitesimal line element centered on the point source, running from $\hat x_1^-$ to $\hat x_1^+$.  This gives a condition on the discontinuity of the field gradient at the particle:
\be
\frac{m}{\mu M^2} \varphi_1 = \varphi'(\hat x_1^+) - \varphi'(\hat x_1^-)~.
\label{deriv-support}
\ee
Differentiating Eq.~\eqref{single-particle-soln}, we find a quadratic expression for $\varphi_1$, leading to
\be
\varphi_1 = \frac{1}{2 \sqrt{2}} \left( \pm \sqrt{ \frac{m^2}{\mu^2 M^4} + 8} - \frac{m}{\mu M^2}  \right)~.
\label{singlepart-value}
\ee
 We take the positive branch so that, as $m \to 0$, we have $\varphi \to 1$ everywhere.

This expression depends only on the dimensionless combination $m / \mu M^2$:\be
\varphi_1 \approx \begin{cases}
1 - \frac{m}{2 \sqrt 2 \mu M^2} & m \ll \mu M^2~, \\
\frac{\sqrt 2 \mu M^2}{m} & m \gg \mu M^2~.
\end{cases}
\label{varphi_0-approx}
\ee
We see that small particle masses do not perturb $\varphi$ very far from the VEV.  On the other hand, large particle masses drive the field towards zero in the vicinity of the particle.

\subsection{Point particle screening}
\label{subsec:SP-screening}

\begin{figure}
\centering
\includegraphics[scale=0.4]{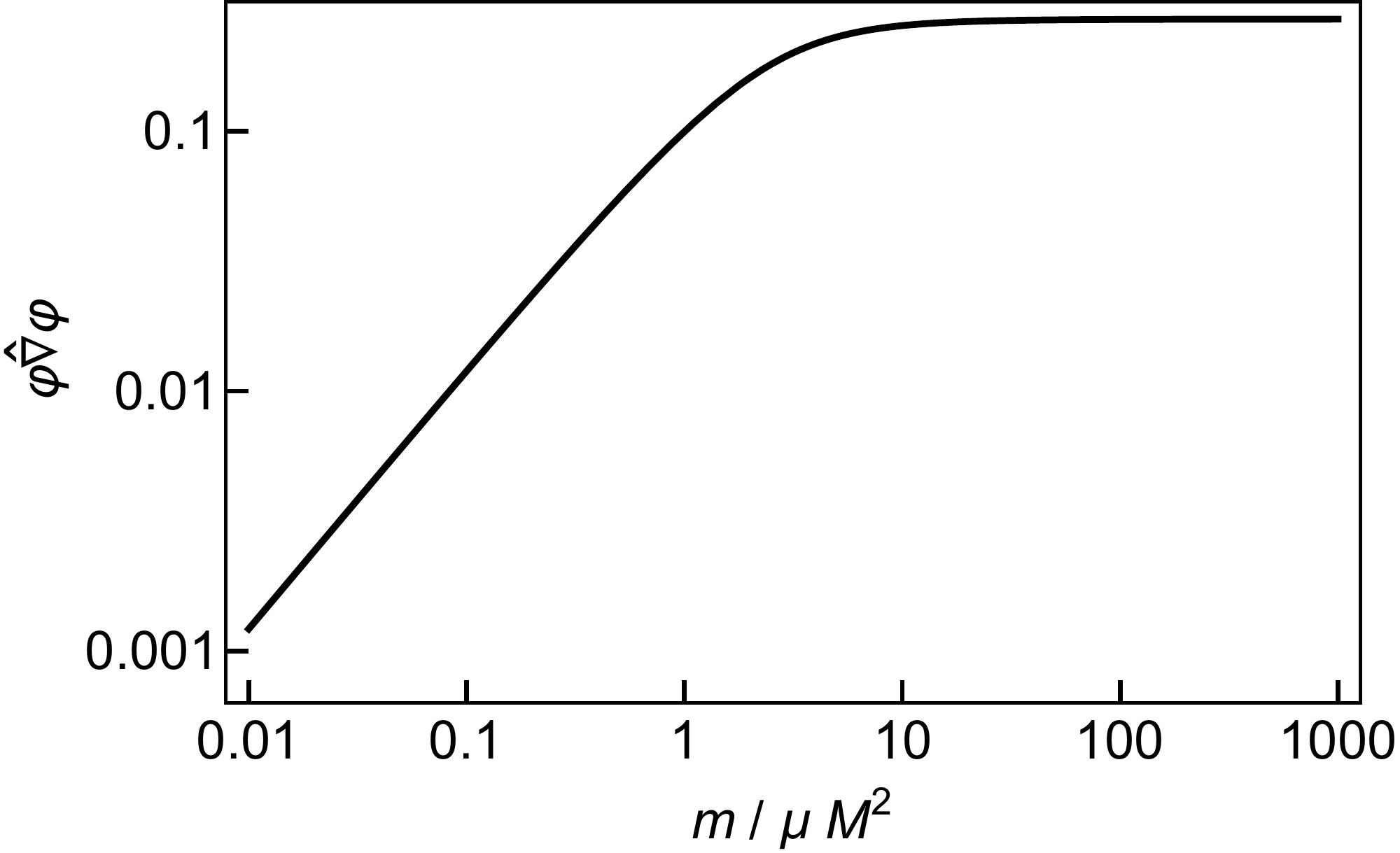}
\caption{\small The magnitude of the attractive force on an unscreened test mass $\varphi \hat \nabla \varphi = \frac{M^2}{\mu \phi_\infty^2} |F|$ due to a point source at the origin, evaluated one Compton wavelength away at $\hat x = 1$.  The field $\varphi$ is obtained from Eq.~\eqref{single-particle-soln}, and the force is computed from Eq.~\eqref{test-force}.   For $m < \mu M^2$, the force grows linearly with $m$ and the source is unscreened.  When $m > \mu M^2$, the force approaches a maximum where further increases in $m$ do not result in a proportionally greater force, i.e. the particle is screened.}
\label{single-particle-force}

\centering
\includegraphics[scale=0.38]{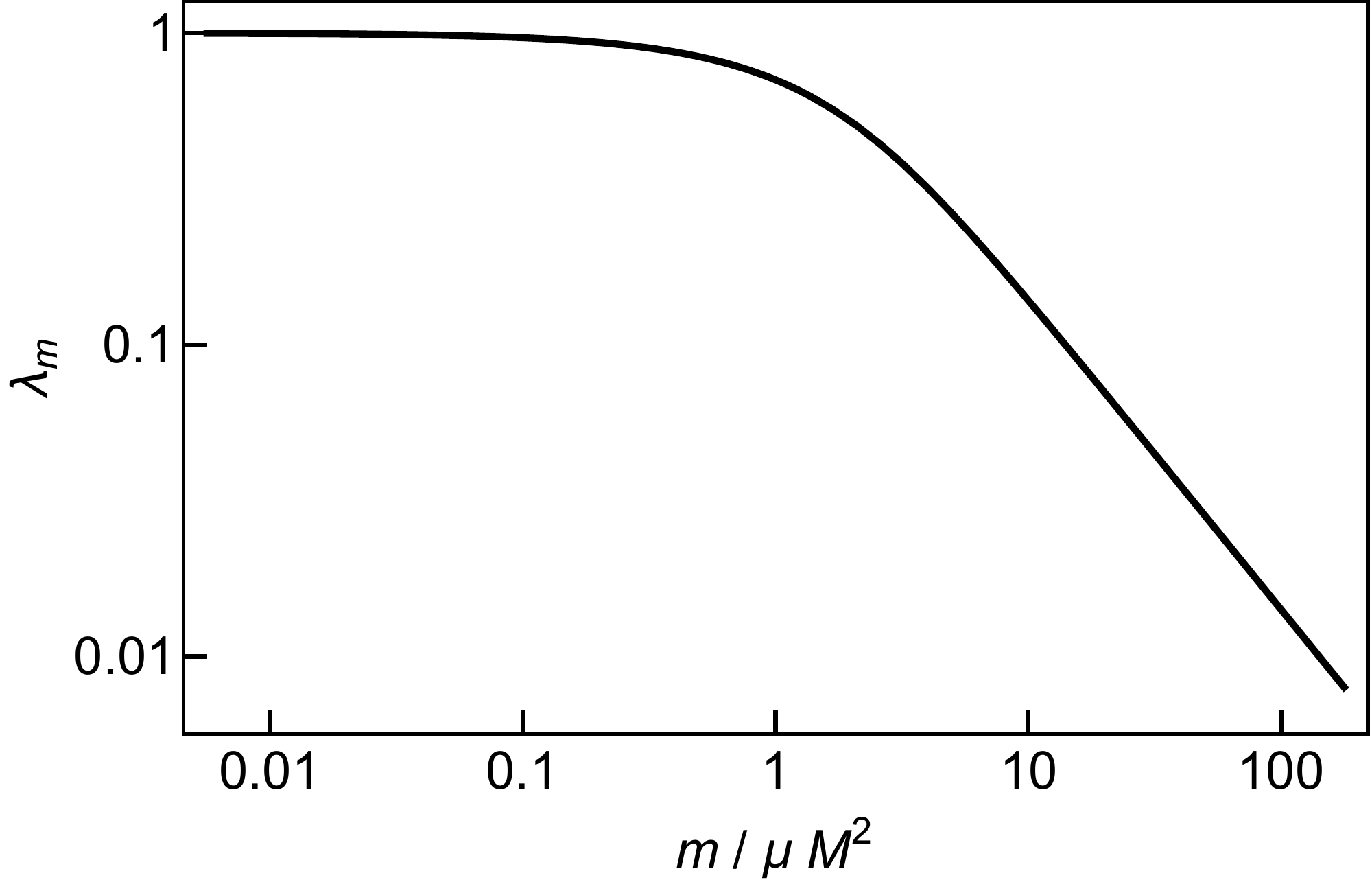}
\caption{\small The screening factor of an isolated point source in vacuum.  The source is screened when its mass exceeds a critical value $m^* = \mu M^2$.  Intuitively, the critical mass $m^*$ is the point at which one could smear out the mass over one Compton wavelength (of the symmetron), and the region would have sufficient density $\rho = \mu^2 M^2$ to remove the false vacuum in the effective symmetron potential, cf.~Eq.~\eqref{full-lagr}.}
\label{single-particle-screening}
\end{figure}

Now that we have an exact solution for the field around a point source we can compute the scalar force that the field exerts on a test mass.  Throughout this subsection we revert to our original conventions for $\phi$ and $x$, as these are the most intuitive variables for computing the force.  In these variables, the solution for the field around a point source is
\be
\phi_\mathrm{point}(x) = \phi_\infty \tanh \left( \frac{\mu}{\sqrt 2} | x - x_1 | + \arctanh \varphi_1   \right)~.
\label{phi-point}
\ee
The force on a unit test mass is given by Eq.~\eqref{test-force}.  It is particularly interesting to note how the scalar force, at a given distance from the source, varies with the mass of the point source.  This relationship is plotted in Fig.~\ref{single-particle-force}.  We see that for small point source masses $m \ll \mu M^2$, the force grows linearly with $m$.  This behavior is consistent with what one expects for an unscreened source: increasing the mass of the source results in a proportionally larger force on the test particle.

However, the curve flattens for larger point source masses $m > \mu M^2$.  In this regime, increasing the mass beyond $m = \mu M^2$ does not result in a proportionally stronger force, which is the hallmark of screening.  Evidently, point particles can be screened, and the amount of screening depends only on the mass of the particle.

We can compute the amount of screening in the following way.  As was done in~\cite{Hui:2009kc}, we imagine that  a small source $A$ with mass $m$ is in the vicinity of another source $B$, and we solve for the force exerted on $A$ by $B$.  All we require of the sources is that $A$ be small enough that the background field due to $B$ is approximately linear in its vicinity.

Once the scalar force is computed, we will note how its magnitude scales with the mass of the sources.  We will then define a {\it  screening factor} that captures all non-linear mass dependence.

If the point particle were unscreened, the force on a unit test mass would be given by Eq.~\eqref{test-force}. Since we expect some form of screening, we define the screening factor $\lambda_m$ here, which ranges between 0 (completely screened) and 1 (unscreened):
\begin{align} \nonumber
F &= - \lambda_m m \frac{ 1}{M^2} \phi_B \nabla \phi_B~, \\
&\approx - \lambda_m m \frac{\phi_\infty}{M^2} \phi_B'(x_A)~,
\label{force-screening}
\end{align}
where in the second line we have assumed that the field due to object $B$ is only a small perturbation from the VEV at $x_A$, the location of the point particle $A$.

We solve for the screening factor by computing the scalar force explicitly.  Following Ref.~\cite{Hui:2009kc}, we compute the momentum of $A$ along the $i$-th direction, in $n$ spatial dimensions, as
\be
P_i = \int {\rm d}^nx T_i^{~0}~,
\ee
where $T_{\mu \nu} = T_{\mu \nu}^\mathrm{m} + T_{\mu \nu}^\varphi$ is the sum of the energy-momentum tensors of the matter and scalar fields.

The volume integral is performed over an $n$-sphere that is small enough that the total energy-momentum inside is dominated by the mass of the object itself, and not by the field content outside the object.  This ensures that the momentum of the sphere is a good representation of the momentum of the object.  Of course, the sphere must also be large enough to encompass the entire object.  Since $A$ is a $\delta$-function in our calculation, this condition is automatically satisfied.

The scalar force on the object is then
\begin{align}
\dot P_i &= \int {\rm d}^nx \partial_0 T_i^{~0} = - \int {\rm d}^nx \partial_j T_i^{~j}~,
\end{align}
where the second equality follows from energy-momentum conservation.
In one spatial dimension this is trivially integrated to give
\be
\dot P = T_x^{~x}(x_1) - T_x^{~x}(x_2)~,
\ee
where $x_1 < x_A < x_2$.  We plug in $T_x^{~x} = \frac{1}{2} \phi'^2 - V(\phi)$ (because the matter fields vanish on the boundary of the sphere), and take the limit $x_1 \to x_A^-$ and $x_2 \to x_A^+$.  We assume that $\phi$ is continuous everywhere, so the potential terms cancel, leaving us with
\be
\dot P = \frac{1}{2} ( \phi'(x_A^-)^2 - \phi'(x_A^+)^2 )~.
\label{Pdot}
\ee
The field $\phi$ is approximately a sum of the background and the point-source solution:
\be
\phi(x) \approx \phi_A(x) + \phi_B(x)~,
\label{obj-bg-split}
\ee
where $\phi_A$ is assumed to be given by the point-source solution Eq.~\eqref{phi-point} and $\phi_B$ is left general.
Plugging this in, we find
\be
F = -  \mu \sqrt{2} \left( 1 - \varphi(x_A)^2 \right) \phi_\infty \phi_B'(x_A)~,
\label{force-explicit}
\ee
Comparing Eqs.~\eqref{force-screening} and \eqref{force-explicit}, we find that the screening factor for the point mass must be
\be
\lambda_m = \frac{\sqrt 2 \mu M^2}{m} (1 - \varphi(x_A)^2)~.
\label{sp-sf}
\ee
This expression is plotted in Fig.~\ref{single-particle-screening}, for $\varphi$ given by Eq.~\eqref{singlepart-value}.  The limiting cases are
\be
\lambda_m \approx \begin{cases}
1 - \frac{1}{2 \sqrt 2} \frac{m}{\mu M^2} & m \ll \mu M^2~, \\
\sqrt 2 \frac{\mu M^2}{m} & m \gg \mu M^2~.
\end{cases}
\ee
We see that for small $m$, the mass of the particle is unscreened: $\lambda_m \approx 1$.  However, for particle masses greater than $\mu M^2$, we find that some of the mass of the particle is screened and $\lambda_m \to 0$.

\subsection{Odd solution}
Recall that we assumed the boundary condition $\varphi \to 1$ as $\hat x \to \pm \infty$.  Thanks to the $\mathbb{Z}_2$ symmetry of the action there are two VEVs, located at $\varphi = \pm 1$.  We could equally well have imposed $\varphi \to -1$ at spatial infinity, which would still lead us to Eq.~\eqref{single-particle-soln} but with a minus sign out front.

We could also have imposed a mixed boundary condition where $\varphi \to \pm 1$ as $\hat x \to \pm \infty$, or vice versa.  Repeating the above procedure with the new boundary condition, we find
\be
\varphi_\mathrm{odd}(\hat x) =  \tanh \left( \frac{1}{\sqrt 2} (\hat x - \hat x_1) \right )~.
\label{single-odd-soln}
\ee
Notice the lack of an integration constant $\varphi_1$: this field configuration is independent of the mass of the source.

One might ask which boundary condition is more physically sensible.  We are interested in the ground state configuration, so we should find the boundary conditions that result in the lowest energy configuration for the field.  The Hamiltonian density is
\begin{align} \nonumber
{\cal H} &= \mu^2 \phi_\infty^2 \left( \frac{1}{2} \varphi'^2 + \frac{1}{2} \left( \frac{\rho}{\mu^2 M^2} - 1 \right) \varphi^2 + \frac{1}{4}  \varphi^4 \right)~, \\
&= \mu^2 \phi_\infty^2 \left( \frac{1}{4} + \frac{1}{2} \left( \frac{\rho}{\mu^2 M^2} - 2 \right) \varphi^2 + \frac{1}{2} \varphi^4 \right) ~,
\end{align}
where in the second line we have rewritten $\varphi'$ via the solutions Eqs.~\eqref{single-particle-soln} and \eqref{single-odd-soln}.
Integrating over all space, the Hamiltonian for the odd solution is
\begin{align} \nonumber
H_\mathrm{odd} &= \frac{1}{\mu}\int_{-\infty}^\infty {\rm d}\hat x {\cal H}_\mathrm{odd}~, \\
&= \mu \phi_\infty^2 \left( - \frac{\hat x}{4} \Big |_{-\infty}^\infty + \frac{2 \sqrt 2}{3} \right)~.
\end{align}
The infinite piece, common to the Hamiltonians for both the even and odd configurations, is present because $V(\varphi) \neq 0$ at the VEV. 

The Hamiltonian for the even solution is
\be
H_\mathrm{even} =  H_\mathrm{odd}
+ \mu \phi_\infty^2 \left( \frac{1}{2} \frac{m}{\mu M^2}\varphi_1^2 - \frac{\sqrt 2}{3} \left( 3 \varphi_1 - \varphi_1^3 \right) \right)~.
\ee
Since $\varphi_1 \in (0, 1)$, and assuming Eq.~\eqref{varphi_0-approx} holds, the combined last two terms are negative and the energy of the even solution is lower than that of the odd solution.   In the limit $\varphi_1 \to 0$ (corresponding to the $m \to \infty$ limit), the two energies are equal.  As $m \to 0, \varphi_1 \to 1$, only the potential energy of the VEV remains: $H_\mathrm{even} =  - \mu^2 \phi_\infty^2 \frac{ \hat x}{4} \big |_{-\infty}^\infty$. Therefore, the even configuration is the ground state as long as $m$ is finite.

\section{Top-hat source}
\label{top-hat}
In this section, we briefly consider the scalar field solution for a source of finite extent in order to show that the point-source solutions we have found agree with the zero-width limit of an extended source. 

 The density profile is taken to be:
\be
\rho = \begin{cases}
\rho_0 & | \hat x| \leq  \hat x_1~, \\
0 & |\hat x| >  \hat x_1~.
\end{cases}
\ee
and we assume that inside the source the density is large enough to restore the symmetry, i.e.~$\rho_0 > \mu^2 M^2$. 
The vacuum solution is the same as before:
\be
\varphi(| \hat x| >  \hat x_1)=  \tanh \left( \frac{1}{\sqrt 2} |\hat x - \hat x_1| + \arctanh \varphi_1 \right)~.
\ee
The calculation for the interior solution is very similar to that of the vacuum solution, Eqs.~\eqref{varphi-eom}--\eqref{single-particle-soln}, so we omit the details for brevity.  The result is
\be
\varphi(|\hat x| < \hat x_1) =  \varphi_0 \nc \left( \hat x \gamma, 1-\frac{ \varphi_0^2}{2\gamma^2} \right)~,
\ee
where $\nc$ is a Jacobi elliptic function and $\gamma \equiv \sqrt{\rho_0 / (\mu^2 M^2) - 1 + \varphi_0^2}$.
The field values at the origin and surface ($\varphi_0$ and $\varphi_1$, respectively) are determined by matching the field and its derivative at  $\hat x_1$.

When the width of the source is small compared to the Compton wavelength of the symmetron field in vacuum, i.e.~$\hat x_1 \ll 1$, the interior solution may be approximated by
\be
\varphi \approx  \varphi_0 \left(1 + \left( \rho_0 / (\mu^2 M^2) - 1 + \varphi_0^2 \right)\frac{\hat x^2}{2} + O(\hat x^4) \right)~.
\ee
At lowest order, the field is constant inside the source so $\varphi_0 \approx \varphi_1$. Matching the derivatives at $\hat x_1$ gives, for a source of total mass $m$,
\be
\varphi_1 \approx \frac{1}{2 \sqrt 2} \left( \sqrt{ \frac{m^2}{\mu^2 M^4} + 8} - \frac{m}{\mu M^2} \right)~.
\ee
This exactly matches our result for a point source, Eq.~\eqref{singlepart-value}.

It is natural to wonder if the results we have found here generalize from 1+1 to 3+1 dimensions.  Unfortunately,   exact solutions in 3+1 dimensions are unknown. However, considering the limit of top-hat density distributions allows us to compare the exact 1+1 solution with what is currently known about solutions in 3+1 dimensions.  

In 3+1 dimensions, an approximate form for the symmetron profile around a static, spherically symmetric extended source of constant density  is typically found by approximating the symmetron potential as quadratic both inside and outside the source, but changing the position of the minimum of the potential and the mass of the field appropriately between the two regimes. As this calculation is discussed in detail in  Ref.~\cite{Hinterbichler:2010es}, we quote here only the key result:  the value of the symmetron field at the surface of a source of radius $R$ and mass $m$ is
\begin{equation}
\varphi(R) = \phi_{\infty}M \sqrt{\frac{4 \pi R}{3 m}}+ \mathcal{O}(R)\;,
\end{equation}
where we have again assumed that the density of the source is sufficiently high to restore the symmetry of the symmetron potential.   (We remind the reader that this expression is for 3+1 dimensions, so $M$ is mass dimension 1.)
As  the width of the source is taken to zero, i.e.~$R\rightarrow 0$, we find that the surface field value goes to zero as $\phi(R) \sim \sqrt R$.  So, similarly to the 1+1 dimensional case, we  expect that the field does not diverge at the position of a point source.  However, unlike the 1+1 dimensional case, this indicates that  the value of the field at the position of the source does not depend on the mass of the source. 

We note, however, that in taking the point-like limit of the extended source in 3+1 dimensions,  the  approximation made  to the form of the symmetron potential breaks down, and the field explores regions of the potential which are not well approximated by a quadratic. Therefore, a full analysis of the 3+1 dimensional case is needed, and we leave this  for future work.

\section{Double identical sources}
\label{sec:double-identical}
Next we solve for the symmetron field around two identical point sources, each of mass $m$, located at $\hat x = \pm \hat x_1$.   As before, a subscript 1 indicates a quantity evaluated at $\hat x_1$, for example $\varphi_1 = \varphi(\hat{x}= \hat x_1 )$.  Likewise, $\varphi_0 \equiv \varphi(\hat x = 0)$.

The solution for $|\hat x| > \hat x_1$ follows from the previous section:
\be
\varphi( |\hat x| > \hat x_1 )= \tanh \left( \frac{1}{\sqrt 2}( |\hat x| - \hat x_1) + \arctanh \varphi_1 \right)~.
\ee
Next we solve for the field between the two charges.  This time we integrate the equation of motion from $\hat x = 0$ to some $\hat x \in (0, \hat x_1)$:
\be
\int_{\varphi(0)}^{\varphi(\hat x)}{\rm d}\varphi\; \frac{\rm d}{{\rm d} \varphi} ( \varphi'^2) = \int_{\varphi(0)}^{\varphi(\hat x)} {\rm d}\varphi\; 2 V,_\varphi~.
\ee
By symmetry $\varphi'(0) = 0$, leaving us with
\be
 \varphi'(\hat x)^2 = 2 \left( V|_{\hat x} - V|_{\hat x \to 0} \right)~.
\ee
Taking the positive root, separating variables and integrating again over the same interval gives
\be
\hat x = \int_{\varphi_0}^{\varphi(\hat x)} \frac{{\rm d} \varphi}{\sqrt{\frac{1}{2} \varphi^4 - \varphi^2 + (\varphi_0^2 - \frac{1}{2} \varphi_0^4)}}~.
\ee
As in Ref.~\cite{Brax:2017hna}, we perform this integral by substituting $y \equiv \varphi / \varphi_0$:
\begin{align} \nonumber
\int_1^y &\frac{{\rm d}y'}{\sqrt{\frac{1}{2} \varphi_0^2 y'^4 - y'^2 + (1 - \frac{1}{2} \varphi_0^2)}} \\
\quad &= \frac{1}{\sqrt{1 - \frac{1}{2} \varphi_0^2}} F \left( \arcsin y, \frac{\varphi_0^2}{2 - \varphi_0^2} \right) \bigg |_1^y~,
\end{align}
where $F(u, b)$ is the elliptic integral of the first kind.
The lower bound of the integral gives the complete elliptic integral of the first kind $K(b) = F(\pi/2, b)$.

Rearranging, we have
\be
F \left( \arcsin y, \frac{\varphi_0^2}{2 - \varphi_0^2} \right) = \hat x  \sqrt{1 - \frac{1}{2} \varphi_0^2} + K \left(\frac{\varphi_0^2}{2 - \varphi_0^2} \right)~.
\ee
The inverse of $F(u, b)$ is the amplitude for the Jacobi elliptic functions, denoted $\am(u, b)$:
\be
y = \sin \left( \am \left( \frac{\hat x}{\sqrt 2} \sqrt{2 -  \varphi_0^2} + K\left(\frac{\varphi_0^2}{2 - \varphi_0^2} \right), \frac{\varphi_0^2}{2 - \varphi_0^2} \right) \right)~.
\ee
The RHS is a Jacobi elliptic function: $\sin( \am( u + K(b), b) ) = \sn(u + K(b), b) = \cd(u, b)$.

Summarizing, we have
\be
 \varphi = \begin{cases}
\varphi_0 \cd \left( \frac{1}{\sqrt 2}|\hat x| \sqrt{2- \varphi_0^2}, \frac{ \varphi_0^2}{2 - \varphi_0^2} \right) & |\hat x| < \hat x_1~, \\
\tanh \left( \frac{1}{\sqrt 2}( |\hat x| - \hat x_1) + \arctanh \varphi_1 \right) & |\hat x| > \hat x_1~.
\end{cases}
\label{double-identical-soln}
\ee

Next we must determine the constants $\varphi_0$ and $\varphi_1$.  The first constraint comes from continuity of $\varphi$ at $\hat x_1$:
\be
\varphi_1 = \varphi_0 \cd \left( \frac{1}{\sqrt 2}\hat x_1 \sqrt{2- \varphi_0^2}, \frac{ \varphi_0^2}{2 - \varphi_0^2} \right)~.
\label{first-condition}
\ee
The second constraint is on the gradient of the field.  Exactly as was done in the single particle case, we integrate the equation of motion from $\hat x_1^-$ to $\hat x_1^+$, finding
\be
\frac{m}{\mu M^2} \varphi_1 = \varphi'(\hat x_1^+) - \varphi'(\hat x_1^-) ~.
\label{deriv-support-x1}
\ee
Differentiating Eq.~\eqref{double-identical-soln}, we obtain
\begin{align} \nonumber
&\frac{m}{\mu M^2} \varphi_1 = \frac{1}{\sqrt 2} ( 1 - \varphi_1^2)
+ \varphi_0 \frac{1 - \varphi_0^2}{\sqrt{1 - \frac{1}{2} \varphi_0^2}} \\
&\times \nd \left( \frac{\hat x_1}{\sqrt 2} \sqrt{2 - \varphi_0^2},  \frac{ \varphi_0^2}{2 - \varphi_0^2} \right)
\sd \left( \frac{\hat x_1}{\sqrt 2} \sqrt{2 - \varphi_0^2},  \frac{\varphi_0^2}{2 - \varphi_0^2} \right)~.
\label{second-condition}
\end{align}

\begin{figure}
\centering
\includegraphics[scale=0.4]{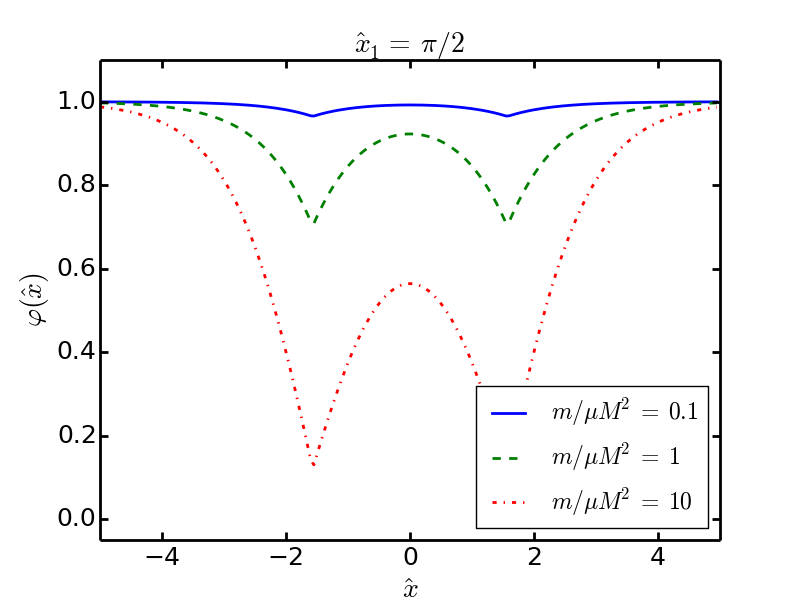}
\includegraphics[scale=0.4]{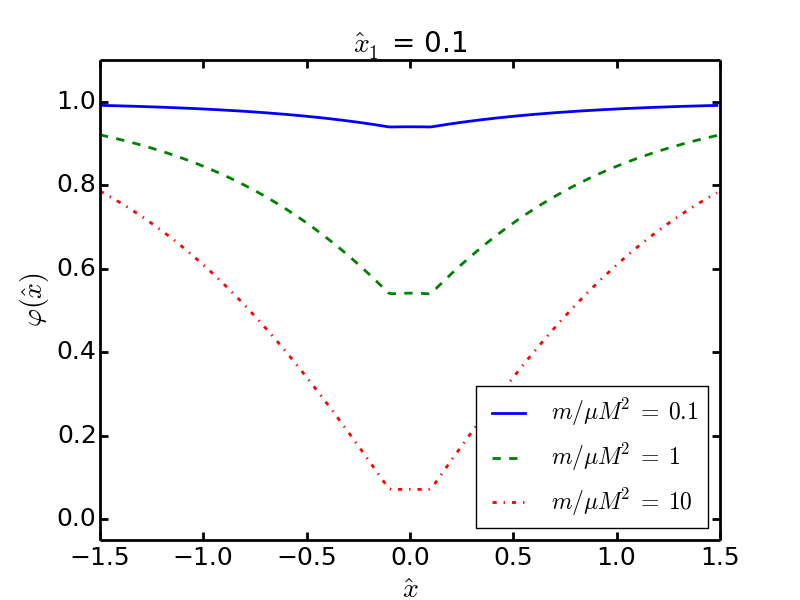}
\caption{\small Field profiles of the two-particle solution, for large particle separation (top) and small particle separation (bottom).  We see that, when the sources are far apart, there is sufficient room for the field to evolve towards the VEV in the middle.  When the particles are close together, the field remains roughly constant in between.  We also see that heavier particle sources pull the field further away from the VEV.  Note also that the field is pushed further from the VEV when the particles are closer together.}
\label{fig-2part-solns}
\end{figure}

No way was found to isolate $\varphi_0$ analytically, so Eqs.~\eqref{first-condition} and \eqref{second-condition} were solved numerically, and the corresponding field profiles are shown in Fig.~\ref{fig-2part-solns}.  We find that for large particle separations $\hat x_1 > 1$, the field is able to recover towards the VEV between the particles.  Conversely, when the particles are close together $\hat x_1 \ll 1$, there is not enough room for the field to recover and the field remains roughly constant between the particles. The rest of this section is devoted to exploring some of the limiting cases of this system.

\subsection{Small separation: $\hat x_1 \ll 1$}

This is the limit in which the particle separation is much smaller than the Compton wavelength of the field.   In this case, the first line of Eq.~\eqref{double-identical-soln} becomes
\be
\varphi(|\hat x| < \hat x_1) = \varphi_0 \left( 1 - \frac{1}{2} (1 - \varphi_0^2) \hat x_1^2  + O(\hat x_1^4) \right)~.
\ee
At lowest order the field is constant between the two sources and we have $\varphi_0 \approx \varphi_1$.

Having established that the gradient approximately vanishes between the sources in this limit, it is straightforward to use Eq.~\eqref{deriv-support-x1} to solve for $\varphi_1$:
\be
\varphi_0 = \varphi_1 = \frac{1}{2 \sqrt 2} \left( \sqrt{ \frac{(2 m)^2}{\mu^2 M^4} + 8} - \frac{2 m}{\mu M^2}  \right)~.
\label{double-near-approx}
\ee
This is the same value as that of a single point source of mass $2m$, given by Eq.~\eqref{varphi_0-approx}.  Evidently two nearby sources, each of mass $m$, depress the symmetron field by the same amount as a single source with total mass $2m$.

We can also compute the screening factor for this system, as was done in Sec.~\ref{subsec:SP-screening}.  The calculation proceeds in the same way as before, except that we integrate over a finite volume $(-\hat x_1^+, \hat x_1^+)$:
\be
\lambda_{2 m} = \frac{\sqrt 2 \mu M^2}{2 m} \left(1 - \varphi_1^2 \right)~.
\ee
As with the field value, we find that the screening factor in this limit is the same as that of a single particle with total mass $2m$.

\subsection{Infinite mass: $\varphi_1 = 0$}

Another useful limit is when the mass of the sources is very large, so that the field vanishes at their locations.  Assuming $\varphi_1 \to 0$ in this limit, as Eqs.~\eqref{varphi_0-approx} and \eqref{double-near-approx} suggest, we find that Eq.~\eqref{first-condition} becomes
\be
0 = \varphi_0 \cd \left( \frac{1}{\sqrt 2}\hat x_1 \sqrt{2- \varphi_0^2}, \frac{\varphi_0^2}{2 - \varphi_0^2} \right)~.
\ee
We use the definition of the Jacobi elliptic function $\cd$ to invert the equation, rearrange and then invert again using the elliptic integral, finding
\be
\frac{1}{\sqrt 2}\hat x_1 \sqrt{2 - \varphi_0^2} = K \left(\frac{ \varphi_0^2}{2 - \varphi_0^2} \right) + F\left( n \pi, \frac{ \varphi_0^2}{2 - \varphi_0^2} \right)~,
\ee
for $n \in \mathbb{Z}$.  We take $n = 0$, leaving us with
\be
\hat x_1 = \frac{1}{\sqrt{1 - \varphi_0^2/2}} K\left(\frac{\varphi_0^2}{2 - \varphi_0^2} \right)~.
\ee
Taylor expanding for small $\varphi_0$, we find
\be
\varphi_0 \approx 4 \sqrt{\frac{1}{3 \pi} \left(\hat x_1 - \pi / 2 \right)}~.
\ee
This expression for $\varphi_0$ becomes imaginary for $\hat x_1 < \pi / 2$, leaving us with the trivial solution $\varphi_0 = 0$ for particle separations less than $\pi$ Compton wavelengths.  This exactly matches the 1D planar solutions previously found in Ref.~\cite{Upadhye:2012rc}. 

The expression for $\hat x_1$ diverges as $\varphi_0 \to 1$, but this is also expected: it just means that the field recovers to the VEV between the particles when the particle separation is large.

\subsection{Large separation: $\hat x_1 \gg 1$}

When the particles are very far apart, i.e.~$\hat x_1 \gg 1$, we expect that the
field configuration around each particle resembles the one we found in the single-source case.  Some care must be taken in this limit, because as $\hat x \to \infty$ we also have $\varphi_0 \to 1$, causing some of the functions in Eq.~\eqref{double-identical-soln} to diverge.

We begin by rewriting the interior solution, via the definition of cd, as:
\be
\varphi(\hat x < \hat x_1) = \varphi_0 \sn \left( \frac{-\hat x}{\sqrt 2} \sqrt{2 - \varphi_0^2} + K \left(\frac{\varphi_0^2}{2 - \varphi_0^2} \right), \frac{\varphi_0^2}{2 - \varphi_0^2} \right)~.
\ee
Now we expand for large $\hat x$, where we also write $\varphi_0 = 1 - \epsilon$, as $\epsilon \to 0$.  In this limit, the parameter of the Jacobi elliptic function also approaches 1, so we expand $\sn(u,b) \approx \tanh(u) + O(1 - b)$.  At lowest order, we have
\be
\varphi(\hat x) =  \tanh \left( \frac{-\hat x}{\sqrt 2} \sqrt{2 - \varphi_0^2} + K \left(\frac{\varphi_0^2}{2 - \varphi_0^2} \right) \right) + O(\epsilon)~,
\ee
where we have not yet applied the expansion to the argument of $\tanh$.  This is justified because $\sn$ and $\tanh$ always evaluate to a number between -1 and 1.

For consistency, we evaluate this expression at $\hat x_1$, which tells us
\be
K\left( \frac{\varphi_0^2}{2 - \varphi_0^2} \right) = \frac{\hat x_1}{\sqrt 2} + \arctanh{\varphi_1} + O(\epsilon)~.
\ee
Substituting into the above expression for $\varphi$, we find that the field profile is approximately a $\tanh$ around each source, just as we found in the single-source case:
\be
\varphi(\hat x) =  \tanh \left( \frac{|\hat x - \hat x_1|}{\sqrt 2} + \arctanh{\varphi_1} \right) + O(\epsilon)~.
\ee

Equation~\eqref{second-condition} allows us to solve for $\varphi_1$.  We can rearrange the elliptic functions $\nd(u, b) \cdot \sd(u, b) = \frac{\sn(u,b)}{1 - b} (1 - b \cd(u,b)^2)$, and then use Eq.~\eqref{first-condition} to give $\cd = \varphi_1 / \varphi_0$. 

With these substitutions, Eq.~\eqref{second-condition} becomes
\begin{align} \nonumber
\frac{m}{\mu M^2} \varphi_1 = \frac{1}{\sqrt 2} ( 1 - &\varphi_1^2)
+ \frac{\varphi_0}{\sqrt 2}
\left( \frac{2 - \varphi_0^2 - \varphi_1^2}{\sqrt{2 - \varphi_0^2}} \right) \\
&\times \sn \left(\frac{\hat x_1}{\sqrt 2} \sqrt{2 - \varphi_0^2}, \frac{\varphi_0^2}{2 - \varphi_0^2}\right)~.
\end{align}
We expand in the double limit $\hat x_1 \gg 1$, $\varphi_0 = 1 - \epsilon$ as $\epsilon \to 0$.  The function $\sn$ again becomes $\tanh$ at lowest order, leading to
\be
\frac{m}{\mu M^2} \varphi_1 = \sqrt 2 ( 1 - \varphi_1^2) + O(\epsilon)~.
\ee
Solving for $\varphi_1$, we have
\be
\varphi_1 = \frac{1}{2 \sqrt{2}} \left( \sqrt{ \frac{m^2}{\mu^2 M^4} + 8} - \frac{m}{\mu M^2}  \right) + O(\epsilon)~,
\label{doubleparticle-largesep}
\ee
which matches Eq.~\eqref{singlepart-value}, viz.~the field value in the single-particle case.  As the distance between the two particles is decreased, $\varphi_1$ smoothly interpolates between the large-separation result and the small-separation one.

\section{The electrostatic analogy}
\label{electrostatics}

One shortcut towards understanding some symmetron phenomenology is via an analogy to electrostatics~\cite{JonesSmith:2011tn,Pourhasan:2011sm}.  The idea is that if the symmetron theory is linearized around the VEV then the (linear) equation of motion for perturbations matches that of electrostatics.  Such an analogy lets us borrow intuition and techniques from a very familiar area of physics.

Expanding the Lagrangian to quadratic order around the VEV, i.e.~writing $\phi = \phi_\infty + \xi$, the equation of motion becomes
\be
(\vec \nabla^2 - m_\xi^2) \xi = \frac{\phi_\infty}{M^2} \rho~,
\label{ES-eom}
\ee
where $m_\xi^2 = V,_{\phi \phi}|_{\phi_\infty}$.  This is the same equation of motion as that of massive electrostatics.  Furthermore, if we restrict our attention to distances much smaller than the Compton wavelength of the field, $ D \ll m_\xi^{-1}$, the mass term may be neglected and our theory for perturbations $\xi$ becomes exactly that of ordinary electrostatics, just with a different coupling constant.  The symmetron force on a unit test mass is now
\be
\vec F = -\frac{\phi_\infty}{M^2} \vec \nabla \xi~.
\ee
A large, dense sphere, taken to be perfectly screened, satisfies $\xi = - \phi_\infty$ inside.  Thus, screened objects are analogous to charged conductors held at a particular equipotential.  The symmetron force between such an object and a point particle may then be computed easily via the method of images, as was shown in Ref.~\cite{Ogden:2017xeo}.

The authors of that paper computed the symmetron force in the electrostatic approximation between a large, perfectly screened sphere and a point source.  Using the method of images, there are two contributions to the force: (1) the image charge of the sphere, which captures the ordinary charge of the sphere, and (2) the image of the test particle, which captures the backreaction of the field due to the presence of the test particle.

They argued that at large distances, (1) dominates and the force is attractive, as expected.  However, as the test particle is brought towards the sphere, (2) dominates and the force becomes repulsive.  This surprising result was taken as an example of how repulsive scalar forces might arise, although they acknowledged that further work is needed to verify this claim.

It is important to keep in mind that the electrostatic analogy is based on a linearization of the theory and must be abandoned when non-linearities become important.  This occurs when the perturbations grow to be of order the VEV: $|\xi| \sim \phi_\infty$.  In the example of the screened sphere, the magnitude of $\xi$ grows as the distance to the sphere is decreased, until eventually the non-linear terms cannot be neglected.

It was in this limit that the apparent repulsive force was found by means of the electrostatics analogy. However, we expect this analogy to break down when the non-linear terms become relevant. In fact, it is possible to see that the force remains attractive under quite general assumptions, and we describe this below for one spatial dimension.

Imagine that we have two objects $m_A$ and $m_B$, located at $x_A$ and $x_B$ respectively.  We take the density of object A to be a delta function $\rho_A = m_A \delta(x_A)$. The scalar force on $A$ due to $B$ was computed in Section~\ref{sec:SP-soln}, where we found
\be
\dot P_i = \frac{1}{2} ( \phi'(x_A^-)^2 - \phi'(x_A^+)^2 )~.
\label{Pdot2}
\ee
The presence of $B$ to one side of $A$ has the general effect of decreasing the field, and its gradient, on that side.  We saw an example of this explicitly in Sec.~\ref{sec:double-identical} (and Fig.~\ref{fig-2part-solns} in particular.) Assuming $x_B < x_A$, we therefore have $\phi'(x_A^-) < \phi'(x_A^+)$ in general.  It then follows from Eq.~\eqref{Pdot2} that $A$ is attracted to $B$.

This argument has been made in one dimension for concreteness, and our example of field gradient suppression by matter is for point particles, but we see no reason to doubt this behavior for collections of particles or continuous distributions of matter in higher numbers of (particularly three) spatial dimensions.  The key behavior that must be true for this argument to work in three spatial dimensions is that the field's gradient must become small in regions of high density.  Since this is consistent with the standard screening behavior of the symmetron, we are optimistic that this result can be generalized to 3D, and we leave the verification of this to future work.

\section{Linear coupling}
\label{sec:linear-coupling}

We now turn our attention to a modification of the symmetron model, wherein the coupling to the density is taken to be linear rather than quadratic. As we will see, this change in the coupling has a marked effect on the screening. Most notably, we will find that there is no single-particle screening and, instead, see evidence of a breakdown when the  mass of the point source exceeds a critical value.

The equation of motion now takes the form
\begin{equation}
\varphi'' = -\varphi(1-\varphi^2)+\frac{1}{\mu^2 M}\rho~.
\end{equation}
For point-like sources, the only impact of this change in the coupling to the density is to modify the boundary conditions.   The vacuum solutions remain the same.

In the case of a single charge, the solution is still of the form in Eq.~\eqref{deriv-support}, but the discontinuity of the field gradient at the point charge becomes
\begin{equation}
\frac{m}{\mu M} = \varphi'(\hat x_1^+)-\varphi'(\hat x_1^-)~,
\end{equation}
yielding
\begin{equation}
\varphi_1 = \pm\sqrt{1-\frac{m}{\sqrt{2}\mu M}}~.
\end{equation}
The boundary condition $\varphi\to 1$ at spatial infinity requires us to take the positive root. We note that the $\mathbb{Z}_2$ transformation $\varphi\to-\,\varphi$, which can be used to absorb any of change of sign in the coupling $m\to -\,m$, cannot be obtained by a straightforward transformation of $\varphi_1$.

We notice that $\varphi_1$ becomes imaginary for $m>\sqrt{2}\mu M$, and we take this branch point in the square-root to signal a breakdown phenomenon. This breakdown results from the fact that the potential is bounded from below at $\varphi=0$, such that the maximum gradient that the field equations can support is bounded from above. Namely, the first integral of the equation of motion implies that
\begin{equation}
    |{\rm disc}\,\varphi_1'| = 2\sqrt{2}\sqrt{V|_{\hat{x} \to  \hat{x}_1}
    -V|_{\hat{x}\to\infty}}~,
\end{equation}
 where $\rm{ disc } ~\varphi_1' \equiv \varphi'(x_1^+) - \varphi'(x_1^-)$.
 It follows that $|{\rm disc}\,\varphi_1'|<\infty$ if $V|_{\hat{x} \to  \hat{x}_1}-V|_{\hat{x}\to\infty} < \infty$. With respect to the dimensionless coordinate, the maximum gradient for the present model is $1/\sqrt{2}$. In contrast, for the quadratic coupling of Sec.~\ref{sec:SP-soln}, the gradient generated by the presence of the source is modulated by the value of the field, and this is precisely the single-particle screening mechanism identified earlier.

The only assumptions which have gone into this derivation are that the solution is static,  that  the field reaches its VEV at spatial infinity, and that the backreaction of the scalar field on the distribution of matter can be neglected. The breakdown we find here may indicate the failure of one or more of these assumptions for high-mass particles.

To illustrate further the role played by the boundedness of the potential at the origin (in field space), we consider the inverse monomial chameleon, whose potential (in 1+1 dimensions) is given by
\begin{equation}
    V(\phi) = \frac{\Lambda^{2}}{\phi^n}~.
\end{equation}
The equation of motion again has a first integral, and the expression for the discontinuity becomes (assuming $\phi \to \infty$ and therefore $V \to 0$ at spatial infinity)
\begin{equation}
    \frac{m}{\Lambda M} = \frac{2\sqrt{2}}{\phi_1^{n/2}}~,
\end{equation}
fixing the value of the field at the origin to be
\begin{equation}
    \phi_1 =  \bigg(\frac{8\Lambda^2M^2}{m^2}\bigg)^{1/n}~.
\end{equation}
In this case, we see that $V|_{x\to x_1}$ is unbounded, and the gradient of the field can be arbitrarily large at the origin, such that there is no breakdown phenomenon.

We note that, for chameleon models with inverse power law potentials $V(\phi) \propto 1/\phi$, analytic solutions are known in 1 + 1 dimension for the form of the field profile between two sources \cite{Burrage:2016lpu,Ivanov:2016rfs}.

\section{Higgs-Yukawa}
\label{sec:higgs}

The linearly coupled symmetron model bears a striking resemblance to the prototype of the Higgs model.

In $1+1$ dimensions, the Higgs-Yukawa fermion Lagrangian takes the form
\begin{equation}
\mathcal{L} \supset -\bar{\psi}i\slashed{\partial}\psi-y\bar{\psi}\phi\psi~,
\end{equation}
where the Yukawa coupling $y$ has mass dimension $1$ and the fermions are mass dimension $1/2$.
This can be recast in terms of dimensionless fields and the dimensionless coordinate $\hat{x}$ as
\begin{equation}
    \hat{\mathcal{L}} \supset -\bar{\hat{\psi}}i\hat{\slashed{\partial}}\hat{\psi}-\frac{yv}{\mu}\bar{\hat{\psi}}\varphi\hat{\psi}~,
\end{equation}
where $\hat{\psi}=\psi/(\mu^{1/2}v)$ and $v \equiv \mu / \sqrt \lambda$. In order to approximate the fermion density by a Dirac delta function, we must assume that the Compton wavelength of the fermion is much smaller than that of the would-be Higgs. If we suppose, therefore, that there exists a fermion whose mass is significantly larger than the Higgs, we can write $\langle \bar{\hat{\psi}}\hat{\psi}\rangle=\delta(\hat x - \hat x_1)$ in the centre-of-mass frame. The equation of motion then becomes
\begin{equation}
\varphi'' = -\varphi(1-\varphi^2)+\frac{y v}{\mu}\delta(\hat x)~,
\end{equation}
having the same solution as the single-charge case with
\begin{equation}
\varphi_1 = \pm\sqrt{1-\frac{y v}{\sqrt{2}\mu}}~.
\end{equation}
We again see the breakdown, this time occurring at  $y_{\rm crit}=\sqrt{2}\mu/v$ (i.e.~when the naive fermion mass $m_{f,{\rm crit}}=y_{\rm crit}v$ is equal to the Higgs mass, i.e. $m_{\varphi}=\sqrt{2}\mu$). If we suppose the fermion obtains its mass from the Yukawa term, as is the case for Dirac fermions of the SM, this would seem to suggest an upper limit on the mass of the fermion. As we try to increase the mass by increasing the strength of the Yukawa coupling, we further suppress the value of the Higgs field in the vicinity of the point particle, until we reach a point where the local value of the Higgs field is zero and the breakdown occurs. Moreover, as we will now show, this mass is parametrically smaller than the Higgs mass $m_{\varphi}$, in contradiction to our original assumption.

The mass arising from the Yukawa coupling is
\begin{equation}
m_f =  y v \varphi_1 = y v\sqrt{1-y v/(\sqrt{2}\mu)} \leq \sqrt{4/27} m_{\varphi}~,
\end{equation}
where the maximum occurs at $y=2\sqrt{2}\mu/(3 v)$. There is also a contribution to the rest mass of the system from the gradient energy in the perturbed Higgs field. This can be estimated from the Hamiltonian of the system, normalized to the vacuum $\varphi=1$:
\begin{equation}
\label{eq:HiggsH}
H = \frac{1}{3}m_{\varphi}(2+\varphi_1)(1-\varphi_1)^2+y v \varphi_1 \leq \frac{2}{3}m_{\varphi}~,
\end{equation}
with the maximum occurring at $y=\sqrt{2}\mu /v$ and reducing to $y v \varphi_1$ in the limit $\varphi_1\to 1$. Notice that, at the maximum, the gradient energy dominates over the vanishing contribution from the Yukawa term.

Given the above results, one might postulate that the observed breakdown bounds the fermion masses attainable from a Yukawa coupling to be of order the Higgs mass itself. Of course, the present analysis was restricted to one spatial dimension.

It will be interesting to see whether similar behavior holds in higher spatial dimensions.  In particular, while exact analytic results cannot be obtained beyond one spatial dimension, it seems plausible that the non-linear terms preclude radially symmetric solutions that diverge, such that the field will still be suppressed in the vicinity of a point source.  Such a finding would be a natural extension of screening behavior seen in the solution to the piecewise-linearized theory Eq.~\eqref{piecewise-linear},  where the field is suppressed at the surface of a dense sphere.

Taking the above results seriously, it is then intriguing that the top-quark mass exceeds that of the Higgs boson in the SM (by roughly a factor of $\sqrt{2}$), and we postpone comprehensive studies of this effect and its potential implications to future work.

\section{Conclusions}
\label{sec:conclusion}
In this paper, we have solved for the symmetron field in one spatial dimension around point particles.  We have seen that, contrary to the conventional understanding of symmetron screening, even a point particle can be screened, provided that its mass is large enough.  We have also seen that pairs of point particles, when separated by less than a Compton wavelength $\mu^{-1}$, behave as a single point particle with the same total mass, and the field is constant in between.

These findings represent a step towards understanding non-linear behavior of the symmetron field in complicated environments.  Our results suggest the following: large extended objects can be modeled as point particles as long as they are smaller than $\mu^{-1}$.  Furthermore, particles in a group behave as individuals if the interparticle separation is greater than $\mu^{-1}$, otherwise, the field is roughly constant between the particles.

Further work, extending these techniques to higher number of dimensions (particularly 3+1) and greater numbers of particles, is needed to verify this general picture.  Such solutions could do much to further reveal the precise nature of symmetron screening.

We have also applied these techniques to a toy Higgs-Yukawa model.  We found that, when the fermionic density is represented as a $\delta$-function source, it is not possible to generate arbitrarily large fermion masses in 1+1 dimensions consistently via a Yukawa coupling to the Higgs field.  As the Higgs-fermion coupling is increased, the local value of the Higgs field decreases, until eventually the Higgs field reaches $0$ and a breakdown occurs.  The largest fermion mass that can be generated would then appear to be of order the Higgs mass itself. In the context of the hierarchy problem, such a result would imply that Dirac mass terms arising from Yukawa couplings are not problematic for the stability of the electroweak scale. Given this intriguing possibility and the fact that the mass of the heaviest particle in the Standard Model --- the top quark --- is of order the Higgs mass, further studies of this behaviour in $3+1$ dimensions and in the fully relativistic regime are warranted. We save such an investigation for future work.

\begin{acknowledgments}
The authors are grateful for helpful conversations with Lasha Berezhiani, Justin Khoury, and Jeremy Sakstein.  We also thank Philippe Brax, Katherine Brown, Harsh Mathur, and Michael Spannowsky, who offered comments on a previous draft of this manuscript. This work was supported by a Leverhulme Trust Research Leadership Award. CB is supported in part by a Royal Society University Research Fellowship.
\end{acknowledgments}



\begin{thebibliography}{99}
 
\bibitem{Englert:1964et} 
  F.~Englert and R.~Brout,
  Phys.\ Rev.\ Lett.\  {\bf 13}, 321 (1964).
  doi:10.1103/PhysRevLett.13.321


\bibitem{Higgs:1964pj} 
  P.~W.~Higgs,
  Phys.\ Rev.\ Lett.\  {\bf 13}, 508 (1964).
  doi:10.1103/PhysRevLett.13.508


\bibitem{Guralnik:1964eu} 
  G.~S.~Guralnik, C.~R.~Hagen and T.~W.~B.~Kibble,
  Phys.\ Rev.\ Lett.\  {\bf 13}, 585 (1964).
  doi:10.1103/PhysRevLett.13.585


\bibitem{Joyce:2014kja} 
  A.~Joyce, B.~Jain, J.~Khoury and M.~Trodden,
  Phys.\ Rept.\  {\bf 568}, 1 (2015)
  doi:10.1016/j.physrep.2014.12.002
  [arXiv:1407.0059 [astro-ph.CO]].


\bibitem{Bertone:2004pz} 
  G.~Bertone, D.~Hooper and J.~Silk,
  Phys.\ Rept.\  {\bf 405}, 279 (2005)
  doi:10.1016/j.physrep.2004.08.031
  [hep-ph/0404175].


\bibitem{Akrami:2018odb} 
  Y.~Akrami {\it et al.} [Planck Collaboration],
  arXiv:1807.06211 [astro-ph.CO].


\bibitem{Adelberger:2003zx} 
  E.~G.~Adelberger, B.~R.~Heckel and A.~E.~Nelson,
  Ann.\ Rev.\ Nucl.\ Part.\ Sci.\  {\bf 53}, 77 (2003)
  doi:10.1146/annurev.nucl.53.041002.110503
  [hep-ph/0307284].


\bibitem{Hinterbichler:2010es} 
  K.~Hinterbichler and J.~Khoury,
  Phys.\ Rev.\ Lett.\  {\bf 104}, 231301 (2010)
  doi:10.1103/PhysRevLett.104.231301
  [arXiv:1001.4525 [hep-th]].


\bibitem{Khoury:2003aq} 
  J.~Khoury and A.~Weltman,
  Phys.\ Rev.\ Lett.\  {\bf 93}, 171104 (2004)
  doi:10.1103/PhysRevLett.93.171104
  [astro-ph/0309300].


\bibitem{Vainshtein:1972sx} 
  A.~I.~Vainshtein,
  Phys.\ Lett.\  {\bf 39B}, 393 (1972).
  doi:10.1016/0370-2693(72)90147-5


\bibitem{Deffayet:2001uk} 
  C.~Deffayet, G.~R.~Dvali, G.~Gabadadze and A.~I.~Vainshtein,
  Phys.\ Rev.\ D {\bf 65}, 044026 (2002)
  doi:10.1103/PhysRevD.65.044026
  [hep-th/0106001].


\bibitem{Nicolis:2008in} 
  A.~Nicolis, R.~Rattazzi and E.~Trincherini,
  Phys.\ Rev.\ D {\bf 79}, 064036 (2009)
  doi:10.1103/PhysRevD.79.064036
  [arXiv:0811.2197 [hep-th]].


\bibitem{JonesSmith:2011tn} 
  K.~Jones-Smith and F.~Ferrer,
  Phys.\ Rev.\ Lett.\  {\bf 108}, 221101 (2012)
  doi:10.1103/PhysRevLett.108.221101
  [arXiv:1105.6085 [astro-ph.CO]].


\bibitem{Pourhasan:2011sm} 
  R.~Pourhasan, N.~Afshordi, R.~B.~Mann and A.~C.~Davis,
  JCAP {\bf 1112}, 005 (2011)
  doi:10.1088/1475-7516/2011/12/005
  [arXiv:1109.0538 [astro-ph.CO]].


\bibitem{Ogden:2017xeo} 
  L.~Ogden, K.~Brown, H.~Mathur and K.~Rovelli,
  Phys.\ Rev.\ D {\bf 96}, no. 12, 124029 (2017)
  doi:10.1103/PhysRevD.96.124029
  [arXiv:1707.06458 [gr-qc]].


\bibitem{Upadhye:2012rc} 
  A.~Upadhye,
  Phys.\ Rev.\ Lett.\  {\bf 110}, no. 3, 031301 (2013)
  doi:10.1103/PhysRevLett.110.031301
  [arXiv:1210.7804 [hep-ph]].


\bibitem{Brax:2017hna} 
  P.~Brax and M.~Pitschmann,
  Phys.\ Rev.\ D {\bf 97}, no. 6, 064015 (2018)
  doi:10.1103/PhysRevD.97.064015
  [arXiv:1712.09852 [gr-qc]].


\bibitem{Nakamura:2018gxf} 
  T.~Nakamura, T.~Ikeda, R.~Saito and C.~M.~Yoo,
  arXiv:1804.05485 [gr-qc].


\bibitem{Bloomfield:2014zfa} 
  J.~K.~Bloomfield, C.~Burrage and A.~C.~Davis,
  Phys.\ Rev.\ D {\bf 91}, no. 8, 083510 (2015)
  doi:10.1103/PhysRevD.91.083510
  [arXiv:1408.4759 [gr-qc]].


\bibitem{Hinterbichler:2011ca} 
  K.~Hinterbichler, J.~Khoury, A.~Levy and A.~Matas,
  Phys.\ Rev.\ D {\bf 84}, 103521 (2011)
  doi:10.1103/PhysRevD.84.103521
  [arXiv:1107.2112 [astro-ph.CO]].


\bibitem{Dehnen:1992rr} 
  H.~Dehnen, H.~Frommert and F.~Ghaboussi,
  Int.\ J.\ Theor.\ Phys.\  {\bf 31}, 109 (1992).
  doi:10.1007/BF00674344


\bibitem{Gessner:1992flm} 
  E.~Gessner,
  Astrophys.\ Space Sci.\  {\bf 196}, no. 1, 29 (1992).
  doi:10.1007/BF00645239


\bibitem{Damour:1994zq} 
  T.~Damour and A.~M.~Polyakov,
  Nucl.\ Phys.\ B {\bf 423}, 532 (1994)
  doi:10.1016/0550-3213(94)90143-0
  [hep-th/9401069].


\bibitem{Pietroni:2005pv} 
  M.~Pietroni,
  Phys.\ Rev.\ D {\bf 72}, 043535 (2005)
  doi:10.1103/PhysRevD.72.043535
  [astro-ph/0505615].


\bibitem{Olive:2007aj} 
  K.~A.~Olive and M.~Pospelov,
  Phys.\ Rev.\ D {\bf 77}, 043524 (2008)
  doi:10.1103/PhysRevD.77.043524
  [arXiv:0709.3825 [hep-ph]].


\bibitem{Brax:2010gi} 
  P.~Brax, C.~van de Bruck, A.~C.~Davis and D.~Shaw,
  Phys.\ Rev.\ D {\bf 82}, 063519 (2010)
  doi:10.1103/PhysRevD.82.063519
  [arXiv:1005.3735 [astro-ph.CO]].


\bibitem{Burrage:2016yjm} 
  C.~Burrage, E.~J.~Copeland and P.~Millington,
  Phys.\ Rev.\ D {\bf 95}, no. 6, 064050 (2017)
  Erratum: [Phys.\ Rev.\ D {\bf 95}, no. 12, 129902 (2017)]
  doi:10.1103/PhysRevD.95.064050, 10.1103/PhysRevD.95.129902
  [arXiv:1610.07529 [astro-ph.CO]].


\bibitem{OHare:2018ayv} 
  C.~A.~J.~O'Hare and C.~Burrage,
  Phys.\ Rev.\ D {\bf 98}, no. 6, 064019 (2018)
  doi:10.1103/PhysRevD.98.064019
  [arXiv:1805.05226 [astro-ph.CO]].


\bibitem{Burrage:2016rkv} 
  C.~Burrage, A.~Kuribayashi-Coleman, J.~Stevenson and B.~Thrussell,
  JCAP {\bf 1612}, 041 (2016)
  doi:10.1088/1475-7516/2016/12/041
  [arXiv:1609.09275 [astro-ph.CO]].


\bibitem{Brax:2016wjk} 
  P.~Brax and A.~C.~Davis,
  Phys.\ Rev.\ D {\bf 94}, no. 10, 104069 (2016)
  doi:10.1103/PhysRevD.94.104069
  [arXiv:1609.09242 [astro-ph.CO]].


\bibitem{Llinares:2018mzl} 
  C.~Llinares and P.~Brax,
  arXiv:1807.06870 [astro-ph.CO].


\bibitem{Brax:2017xho} 
  P.~Brax, S.~Fichet and G.~Pignol,
  Phys.\ Rev.\ D {\bf 97}, no. 11, 115034 (2018)
  doi:10.1103/PhysRevD.97.115034
  [arXiv:1710.00850 [hep-ph]].


\bibitem{Brax:2018zfb} 
  P.~Brax, A.~C.~Davis, B.~Elder and L.~K.~Wong,
  Phys.\ Rev.\ D {\bf 97}, no. 8, 084050 (2018)
  doi:10.1103/PhysRevD.97.084050
  [arXiv:1802.05545 [hep-ph]].


\bibitem{Burrage:2017qrf} 
  C.~Burrage and J.~Sakstein,
  Living Rev.\ Rel.\  {\bf 21}, no. 1, 1 (2018)
  doi:10.1007/s41114-018-0011-x
  [arXiv:1709.09071 [astro-ph.CO]].


\bibitem{Coleman:1973jx} 
  S.~R.~Coleman and E.~J.~Weinberg,
  Phys.\ Rev.\ D {\bf 7}, 1888 (1973).
  doi:10.1103/PhysRevD.7.1888


\bibitem{Burrage:2016xzz} 
  C.~Burrage, E.~J.~Copeland and P.~Millington,
  Phys.\ Rev.\ Lett.\  {\bf 117}, no. 21, 211102 (2016)
  doi:10.1103/PhysRevLett.117.211102
  [arXiv:1604.06051 [gr-qc]].


\bibitem{Hui:2009kc} 
  L.~Hui, A.~Nicolis and C.W.~Stubbs,
  Phys.\ Rev.\ D {\bf 80}, 104002 (2009)
  doi:10.1103/PhysRevD.80.104002
  [arXiv:0905.2966 [astro-ph.CO]].


\bibitem{Burrage:2016lpu} 
  C.~Burrage, E.~J.~Copeland and J.~A.~Stevenson,
  JCAP {\bf 1608}, no. 08, 070 (2016)
  doi:10.1088/1475-7516/2016/08/070
  [arXiv:1604.00342 [astro-ph.CO]].


\bibitem{Ivanov:2016rfs} 
  A.~N.~Ivanov, G.~Cronenberg, R.~Hollwieser, M.~Pitschmann, Jenke T., M.~Wellenzohn and H.~Abele,
  Phys.\ Rev.\ D {\bf 94}, no. 8, 085005 (2016)
  doi:10.1103/PhysRevD.94.085005
  [arXiv:1606.06867 [gr-qc]].
  
  
\end{thebibliography}
\end{document}